\documentclass[12pt]{article}
\usepackage{latexsym}
\usepackage{graphicx}
\usepackage[usenames,dvipsnames]{color}

\def\double{\baselineskip 24pt \lineskip 10pt}
\def\be{\begin{equation}}
\def\ee{\end{equation}} 
\def\bea{\begin{eqnarray}}
\def\eea{\end{eqnarray}}
\def\pa{\partial}
\def\sst{\scriptscriptstyle}
\def\mco{\multicolumn}
\def\CPbar{\hbox{{\rm CP}\hskip-1.80em{/}}}
\def\bu{$\bullet$}
\def\l{\label}
\def\r{\ref}
\def\fn{\footnote}
\def\vc{V^{\frac{2}{3}}}
\def\e{\emph}
\def\rmsd{$\ell_{\textrm{\tiny{rms}}}$ }
\def\rmsdn{$\ell_{\textrm{\tiny{rms}}}$}
\def\nb{$N$-body problem }
\def\nbn{$N$-body problem}
\def\rnb{relational $N$-body problem }
\def\mb{$N$-body }
\def\mbn{$N$-body}

\def\Ra{\Rightarrow}
\def\half{\frac{1}{2}}
\def\nn{\nonumber}
\def\co{\rm{c}}
\def\si{\rm{s}}
\def\wt{\widetilde}
\def\case#1/#2{\textstyle\frac{#1}{#2}}
\def\doublespace{\baselineskip=20pt plus 3pt\message{double space}}
\def\L{\left}
\def\R{right}
\def\noi{\noindent}
\def\vs{\vspace{.2in}}
\def\e{\emph}
\def\noi{\noindent}
\def\id{, i.e., }
\def\p{$\Psi$~}
\def\pn{$\Psi$}

\def\s{${\mathcal S}$ }
\def\sn{${\mathcal S}$}
\def\elb{$I_\textrm{\scriptsize cm}$ }
\def\elbn{$I_\textrm{\scriptsize cm}$}
\def\ccm{$I_\textrm{\scriptsize cm}=0$ }
\def\ccmn{$I_\textrm{\scriptsize cm}=0$}
\def\ecm{$E_\textrm{\scriptsize cm}=0$ }
\def\ecmn{$E_\textrm{\scriptsize cm}=0$}
\def\ec{$E_\textrm{\scriptsize cm}$ }
\def\ecn{$E_\textrm{\scriptsize cm}$}
\def\lcm{${\bf L}_\textrm{\scriptsize cm}=0$ }
\def\lcmn{${\bf L}_\textrm{\scriptsize cm}=0$}
\def\tbe{three-body~}
\def\tben{three-body}
\def\tb{three-body problem }
\def\tbn{three-body problem}
\def\al{Alpha }
\def\aln{Alpha}
\def\pa{Eq.~(\ref{eqm}) }
\def\pan{Eq.~(\ref{eqm})}
\def\el{$\ell_\textrm{\scriptsize rms}$ }
\def\eln{$\ell_\textrm{\scriptsize rms}$}
\def\sc{Schr\"odinger equation }
\def\scn{Schr\"odinger equation}
\def\np{$V_\textrm{\scriptsize Shape}$ }
\def\npn{$V_\textrm{\scriptsize Shape}$}
\def\c{$C$ }
\def\cn{$C$}
\def\isa{$V_\textrm{\scriptsize New}$ }
\def\isan{$V_\textrm{\scriptsize New}$}
\def\po{Poincar\'e }
\def\pon{Poincar\'e}
\def\n{Newtonian }
\def\nn{Newtonian}

\begin{document}



\bibliographystyle{unstr}


\def\double{\baselineskip 24pt \lineskip 10pt}
\textheight 8.9 in \textwidth 6.5 in \oddsidemargin -10pt \topmargin
-30pt

\def\be{\begin{equation}}
\def\ee{\end{equation}} 
\def\bea{\begin{eqnarray}}
\def\eea{\end{eqnarray}}
\def\pa{\partial}
\def\sst{\scriptscriptstyle}
\def\mco{\multicolumn}
\def\CPbar{\hbox{{\rm CP}\hskip-1.80em{/}}}
\def\bu{$\bullet$}
\def\l{\label}
\def\r{\ref}
\def\fn{\footnote}
\def\vc{V^{\frac{2}{3}}}
\def\e{\emph}
\def\rmsd{$\ell_{\textrm{\tiny{rms}}}$ }
\def\rmsdn{$\ell_{\textrm{\tiny{rms}}}$}
\def\nb{$N$-body problem }
\def\nbn{$N$-body problem}
\def\rnb{relational $N$-body problem }
\def\mb{$N$-body }
\def\mbn{$N$-body}


\def\Ra{\Rightarrow}
\def\half{\frac{1}{2}}
\def\nn{\nonumber}
\def\co{\rm{c}}
\def\si{\rm{s}}
\def\wt{\widetilde}
\def\case#1/#2{\textstyle\frac{#1}{#2}}
\def\doublespace{\baselineskip=20pt plus 3pt\message{double space}}
\def\L{\left}
\def\R{right}
\def\noi{\noindent}
\def\vs{\vspace{.2in}}
\def\e{\emph}
\def\noi{\noindent}
\def\id{, i.e., }
\def\p{$\Psi$~}
\def\pn{$\Psi$}
\def\s{${\mathcal S}$ }
\def\sn{${\mathcal S}$}
\def\elb{$I_\textrm{\scriptsize cm}$ }
\def\elbn{$I_\textrm{\scriptsize cm}$}
\def\ccm{$I_\textrm{\scriptsize cm}=0$ }
\def\ccmn{$I_\textrm{\scriptsize cm}=0$}
\def\ecm{$E_\textrm{\scriptsize cm}=0$ }
\def\ecmn{$E_\textrm{\scriptsize cm}=0$}
\def\ec{$E_\textrm{\scriptsize cm}$ }
\def\ecn{$E_\textrm{\scriptsize cm}$}
\def\lcm{${\bf L}_\textrm{\scriptsize cm}=0$ }
\def\lcmn{${\bf L}_\textrm{\scriptsize cm}=0$}
\def\tbe{three-body~}
\def\tben{three-body}
\def\tb{three-body problem }
\def\tbn{three-body problem}
\def\al{Alpha }
\def\aln{Alpha}
\def\pa{Eq.~(\ref{eqm}) }
\def\pan{Eq.~(\ref{eqm})}
\def\el{$\ell_\textrm{\scriptsize rms}$ }
\def\eln{$\ell_\textrm{\scriptsize rms}$}
\def\sc{Schr\"odinger equation }
\def\scn{Schr\"odinger equation}
\def\np{$V_\textrm{\scriptsize Shape}$ }
\def\npn{$V_\textrm{\scriptsize Shape}$}
\def\c{$C$ }
\def\cn{$C$}
\def\isa{$V_\textrm{\scriptsize New}$ }
\def\isan{$V_\textrm{\scriptsize New}$}
\def\po{Poincar\'e }
\def\pon{Poincar\'e}
\def\n{Newtonian }
\def\nn{Newtonian}


\begin{center}

{\bf \Huge{Entropy and Cosmological\\ Arrows of Time}}

\vspace{.3in}

{\bf \Large{Julian Barbour\footnote{Email: julian.barbour@physics.ox.ac.uk.}}}

\vspace{.12in}

College Farm, The Town, South Newington, Banbury, OX15 4JG, UK.

\end{center}

\noindent {\bf Abstract.} Deutsch and Aguirre have recently shown that the solutions of certain dynamical systems typically contain a point of minimum size that they identify as an entropy minimum and from which the size and entropy increase to infinity in both directions of time. They argue that in such systems entropic arrows of time exist without the need for a special condition imposed in the past. In this paper I sharpen and extend the conditions under which such solutions exist but argue that the resulting arrows of time should not be interpreted as entropic since they point towards greater order and not disorder.

\section{Introduction}

In \cite{dag}, Deutsch and Aguirre draw attention to the difficulty of posing conventional Cauchy initial conditions in cosmology and, guided by the principle of least action, explore the possibility of prescribing configurations of finitely many particles at two different times. They find evidence that for positive potentials with $1/r^3$ and $1/(r_0^2+r^2)^{3/2}$ ($r_0$ constant) distance dependence the resulting solutions typically have a minimum size between the two prescribed configurations and that the size increases to infinity in both time directions. Such solutions were shown to exist for the (negative) $1/r$ Newton potential in \cite{bkl} and are called Janus-point solutions in \cite{entaxy,jan}, the point of minimum size being the Janus point. In this note I point out that size of a configuration is problematic if it is taken to characterise the instantaneous state of a universe since it presupposes an unobservable extrinsic absolute scale. This observation does not affect the main conclusion drawn in \cite{dag}---that arrows of time arise naturally without imposition of a special condition in the past---but does, I argue, alter the interpretation of the result. I also point out that the method used to obtain the bidirectional arrows is, in one sense, not as general as is assumed in \cite{dag} but in another is more general.

\section{Janus Points and Bidirectional Arrows\label{bda}}

This part of my note will use the standand Newtonian representation, from which I will then remove degrees of freedom that are redundant if the system considered is, as in \cite{dag}, taken to model an `island universe'. Let the total centre-of-mass energy of the system be
\be
E_\textrm{\scriptsize cm}=\sum_{a=1}^N {{\bf p}^a\cdot{\bf p}^a\over 2m_a}+V,\label{en}
\ee
where the potential $V$ is a function of the inter-particle separations $r_{ab}=|{\bf r}_a^{\,\textrm{\scriptsize cm}}-{\bf r}_b^{\,\textrm{\scriptsize cm}}|$ and is homogeneous of some degree $k$; here and below ${\bf r}_a^{\,\textrm{\scriptsize cm}}$ is the centre-of-mass position of particle $a$ and the momentum of particle $a$ is ${\bf p}_a=m_a\dot{\bf r}_a^{\,\textrm{\scriptsize cm}}$, where the dot denotes the derivative wrt the time. The centre-of-mass moment of inertia (half the trace of the inertia tensor) measures the size of the system and is
\be
I_\textrm{\scriptsize cm}=\sum_{a=1}^N m_a\,{\bf r}_a^{\,\textrm{\scriptsize cm}}\cdot{\bf r}_a^{\,\textrm{\scriptsize cm}}.\label{icm}
\ee
Its first time derivative is
\be
\dot I_\textrm{\scriptsize cm}=2\sum_{a=1}^N m_a\,\dot{\bf r}_a^{\,\textrm{\scriptsize cm}}\cdot{\bf r}_a^{\,\textrm{\scriptsize cm}}=2\sum_{a=1}^N{\bf p}_a^{\,\textrm{\scriptsize cm}}\cdot{\bf r}_a^{\,\textrm{\scriptsize cm}}=2D,\label{dil}
\ee
where $D$ has the same dimensions as angular momentum and, being a measure of overall expansion, may, as it first was in \cite{2003}, be called the \e{dilatational momentum}. Whereas angular momentum is conserved if the potential is rotationally invariant, $D$ is only conserved if $V$ is homogeneous of degree $-2$. This is most directly seen in a theory with the reparametrisation-invariant geodesic action
\be
A={1\over 2}\int\textrm d\lambda\sqrt{V\left({\textrm d{\bf r}_a^{\,\textrm{\scriptsize cm}}\over\textrm d\lambda}\cdot{\textrm d{\bf r}_a^{\,\textrm{\scriptsize cm}}\over\textrm d\lambda}\right)},\label{geo}
\ee
in which $\lambda$ is an arbitrary monotonic lable and $A$ is invariant under scaling only if $V$ is homogeneous of degree $-2$ to balance the [length]$^2$ dimension of the kinetic term.

After this observation, we obtain by differentiation of (\ref{dil})
$$
\ddot I_\textrm{\scriptsize cm}=2\sum_{a=1}^N m_a\,\ddot{\bf r}_a^{\,\textrm{\scriptsize cm}}\cdot{\bf r}_a^{\,\textrm{\scriptsize cm}}+2\sum_{a=1}^N m_a\,\dot{\bf r}_a^{\,\textrm{\scriptsize cm}}\cdot\dot{\bf r}_a^{\,\textrm{\scriptsize cm}},\label{int}
$$
in which the second term is simply four times the kinetic energy $T$ and the first can be transformed to
$$
-2\sum_{a=1}^N m_a\,{\partial V\over \partial {\bf r}_a}\cdot{\bf r}_a^{\,\textrm{\scriptsize cm}}\equiv -2kV,
$$
in which Newton's second law yields the first expression and Euler's law for homogeneous functions the second. Therefore
$$
\ddot I_\textrm{\scriptsize cm}=-2kV+4T\equiv 4(T+V)-2kV-4V=4E-2(2+k)V,
$$
or
\be
\ddot I_\textrm{\scriptsize cm}=4E-2(2+k)V,\label{lj}
\ee
which in \mb theory is called the Lagrange--Jacobi relation; for secularly confined systems it leads to the virial theorem \cite{eg}, a fact that will be important in the discussion later. If the rhs of (\ref{lj}) is positive definite and $I_\textrm{\scriptsize cm}$, which cannot be negative, does not vanish (an exceptional case I will discuss later), then the graph of $I_\textrm{\scriptsize cm}$ as a function of the time is concave upward with a unique Janus point of minimum size at which $D=0$. Note that $D$ varies in the range $(-\infty,\infty)$ for both directions of Newtonian time and is a Lyapunov variable.

If now the conserved energy $E$ is non-negative, then $\ddot I_\textrm{\scriptsize cm}$ will be positive definite if
\be
X=-2(2+k)V>0.\label{x}
\ee
To obtain conclusions valid at all times, we need $V>0, V=0$ or $V<0$. The corresponding possibilities are:

\vspace{.1in}

\begin{enumerate}

\item If $V>0$ and $k<-2$, then $X>0$ and all solutions have a Janus point provided $I_\textrm{\scriptsize cm}$ does not vanish. This includes the example of \cite{dag} for which $k=-3$ and therefore Janus points are generic. 

\item The simplest case with $V=0$ corresponds to inertial motion with purely kinetic and necessarily positive energy. Without using (\ref{lj}), Carroll \cite{car} recognised around 2004 that inertial motion leads to bidirectional arrows of time. He and Guth have long discussed it (see the online lecture \cite{jo}) but without publishing the example.

\item When $V<0$, we have $\ddot I_\textrm{\scriptsize cm}>0$ if $k>-2$. This case includes the Newton potential, which is negative definite and has $k=-1$. It is the subject of \cite{bkl}, has many interesting aspects \cite{entaxy,jan}, and will be discussed further below.

\item The limiting case $V<0, k=-2$ is also interesting. For it $D$ is a constant of the motion and can be positive, negative or zero. If $D>0$, we have an explosion from zero size and if $D<0$ a collapse to zero size; I am not aware of study of these possibilities. If $D=0$ and $E_\textrm{\scriptsize cm}=0$, the size of the system, as measured by $I_\textrm{\scriptsize cm}$, remains constant and, as shown in \cite{2003}, geodesic motion (governed by an action principle of the form (\ref{geo})) is realised in the space of possible shapes of the system.

\item Finally, there exist exceptionally interesting zero-measure \mb solutions---with the $k=-1$ Newton potential---in which $I_\textrm{\scriptsize cm}=0$ at some instant. Such solutions are said to terminate in total collisions or, by time-reversal symmetry, begin with total explosions \cite{jan}, chapter 16, for which, in agreement with (\ref{lj}), $I_\textrm{\scriptsize cm}$ increases from $0$ to $\infty$ (there is only `half' of a Janus-point solution).

\vspace{.1in}

\end{enumerate}

The conclusions drawn above can now be compared with the results reported in \cite{dag}. It is clear that, as employed in \cite{dag}, the principle of least action must yield genuine solutions.  In accordance with case 1 above, the specific $C/r^3$ potential (with $C$ a positive constant) studied in \cite{dag} must have generic Janus-point solutions provided the energy is non-negative, which it necessarily is since $T$ is always positive and so too is the potential by the assumption $C>0$. However, the only actual calculations reported in \cite{dag} assumed configurations with the same $I_\textrm{\scriptsize cm}$ at the limits of the action integral. Since Janus-point solutions are generic for the studied potential, two different configurations with the same $I_\textrm{\scriptsize cm}$ must necessarily have a unique Janus point between them. However, if two configurations with different $I_\textrm{\scriptsize cm}$ are prescribed, there is no guarantee the Janus point, which must exist, will lie between them since both configurations could be on the same side of it. In the introduction I said the method used in \cite{dag} is, in one sense, not as general as is assumed in \cite{dag} but in another is more general. It is not as general because finding a Janus point is only guaranteed for equal values of $I_\textrm{\scriptsize cm}$, not necessarily for unequal values. It is more general because Janus-point solutions exist for all the cases 1--4 above and not only the specific $1/r^3$ potential studied in \cite{dag}. As regards the `soft' potential $C/(r_0^2+r^2)^{3/2}$ with $C>0$ considered in \cite{dag}, it reduces to $C/r^3$ as $r_0^2\rightarrow 0$, for which by case 1 Janus-point solutions certainly exist and although we can no longer use the Lagrange---Jacobi relation (\ref{lj}) (since the modified potential is no longer homogeneous) it seems possible similar solutions will exist for $r_0\ne 0$.

In the second part of this note I now consider the interpretation of Janus-point solutions in order to argue that their bidirectional arrows of time should not be interpreted as entropic. This will require some preliminary observations; more details can be found in \cite{jan}.

\section{The Effect of Newton's Absolutes in Dynamics\label{abs}}

When he created dynamics, Newton introduced five absolutes: time, its direction, position and direction in space, and scale. None of these is needed to represent the irreducible content of dynamics, which solely concerns the relations between objects (mass ratios of the particles and ratios of the separations between them in the \nbn); indeed, the `absolutes' introduce redundant gauge degrees of freedom (dofs) into the \nb if, as often, it is used to model an `island universe' in Euclidean space.\footnote{An inertial frame of reference is widely employed to describe subsystems \e{within} the universe. It retains the `absolutes' but harmlessly: they express the relationship between it and the universe at large. But the universe is all that is. For it I will argue the `absolutes' can lead to questionable conclusions.\label{if}} They can be eliminated and the true physical content of an \mb universe established in an illuminating manner by representing conventionally determined \mb solutions in \e{shape space} \sn, which is obtained from the standard Newtonian configuration space by quotienting by Euclidean translations, rotations, and dilatations. The inclusion of dilatations means that one quotients by the similarity group and not simply the Euclidean group. This difference is critical.

Any solution of the \nbn, to which I restrict myself, can, after being found using Newton's laws in an inertial frame, be projected to \s as a mere succession of shapes. It becomes an \e{undirected and unparametrized} curve in \sn. None of the physical information is lost---it is implicitly encoded in the bending of the curve in \s and above all in the shapes through which it passes. The factors that govern the bending are identfied in \cite{jan}, chapter 8, and will be listed here. Note first that one can, using the geodesic action (\ref{geo}) in the Newtonian configuration space, project the resulting geodesics to \sn, where the Cauchy data, as the relevant dofs for one consist of a point in \s and a direction in \s at it. In contrast, Newtonian solutions when projected to \s need more Cauchy data than that because there are five dimensionless global Newtonian quantities included in \n initial data that cannot be encoded in a shape and a direction.

To identify them, note first that in accordance with the velocity decomposition theorem \cite{rings} the kinetic energy $T$ can be decomposed at any instant into three orthogonal components $T_\textrm{\scriptsize s}, T_\textrm{\scriptsize r},T_\textrm{\scriptsize d}$, which are respectively the kinetic energies in change of shape, rotation, and overall expansion ($d$ standing for dilatation). From these three dimensionful quantities one can form the two dimensionless ratios $T_\textrm{\scriptsize r}/T_\textrm{\scriptsize s}$ and $a=T_\textrm{\scriptsize d}/ T_\textrm{\scriptsize s}$, the designation of which by $a$ will be explained shortly. Two more numbers determine the instantaneous direction of the angular momentum within the current shape and one final fifth pure number is the ratio $T/V$ of the two components of the energy $E$. If $E=T+V=0$, then $T/V=-1$ at all times; otherwise it varies in time. The existence of these five pure numbers means that the Cauchy data for general \mb solutions in \s include besides the shape and direction needed for a geodesic the five extra data just listed. In a relational Machian theory \cite{jan} the energy and angular momentum vanish permanently, so for Machian solutions the only remaining additional Cauchy datum is
\be
a=T_\textrm{\scriptsize d}/ T_\textrm{\scriptsize s}.\label{ano}
\ee
It varies along the solution curve, determining the amount it bends away from any reference geodesic tangent to it at the given shape. I call $a$ the \e{anomaly} because there is nothing in the relational notion of shape space to expect it to play a role. It appears in solutions of the \nb projected to \s because the Newton potential is homogeneous of degree $k=-1$ and not $k=-2$, for which, as noted above with the action (\ref{geo}), geodesic theories, the ones that are maximally predictive in \sn, exist. I will say more about (\ref{ano}) below.

\section{The Sense in which the Universe Expands\label{ue}}

It is obvious that our universe is not observed to be expanding relative to an absolute external ruler. Quite the contrary, \e{ratios} of quantities that can be observed within the universe are changing. One is the typical diameter of galaxies divided by the typical intergalactic separation; it is manifestly decreasing. If there is one, the justification for saying the universe is expanding is that during cosmological evolution the ratios of the sizes of many different entities---atoms, molecules, planets, stars, and galaxies---stay more or less constant while the separations between the galaxies increases. All statements about expansion of the universe reduce to variations of such ratios, which are \e{pure numbers}. 
Epistemologically, that is, in terms of what can actually be observed, the universe since the big bang has not been expanding but changing its shape, this being expressed through the emergence of subsystems---the entities just listed---that after their creation serve as effective rods by maintaining with greater or lesser accuracy constancy of the ratios of their sizes while decreasing in size compared with the typical intergalactic separations. Any satisfactory model of the universe must include emergent rods and also clocks, which together are critical in establishing all known facts about its expansion.

It is of course recognised in cosmology that it is meaningless to say the universe has a given size at a certain instant but it is assumed that ratios of sizes at different instants are meaningful. However, the simplest homogeneous and isotropic FLRW cosmologies remain structureless at all times and are therefore flawed models of the actual universe. Models that are realistic in being inhomogeneous are obtained by perturbing FLRW models, but they are still problematic. First, the perturbed models retain lengths and times without models of clocks and rods that measure them. Indeed, Einstein himself admitted \cite{auto} to committing a `sin' in his creation of general relativity as an incomplete theory that postulates the existence of two critical but unrelated entities: on the one hand, the spacetime metric and, on the other, rods and clocks `brought in' to measure the metric rather than emerging from the fundamental equations of the theory. Thus, even when perturbed, the FLRW models retain the `original sin'. The second problem with the perturbed FLRW solutions is that the difficulty of defining simultaneity in general relativity introduces gauge uncertainty into the models when they are perturbed. 

\begin{figure}
\begin{center}
\includegraphics[width=0.9\textwidth]{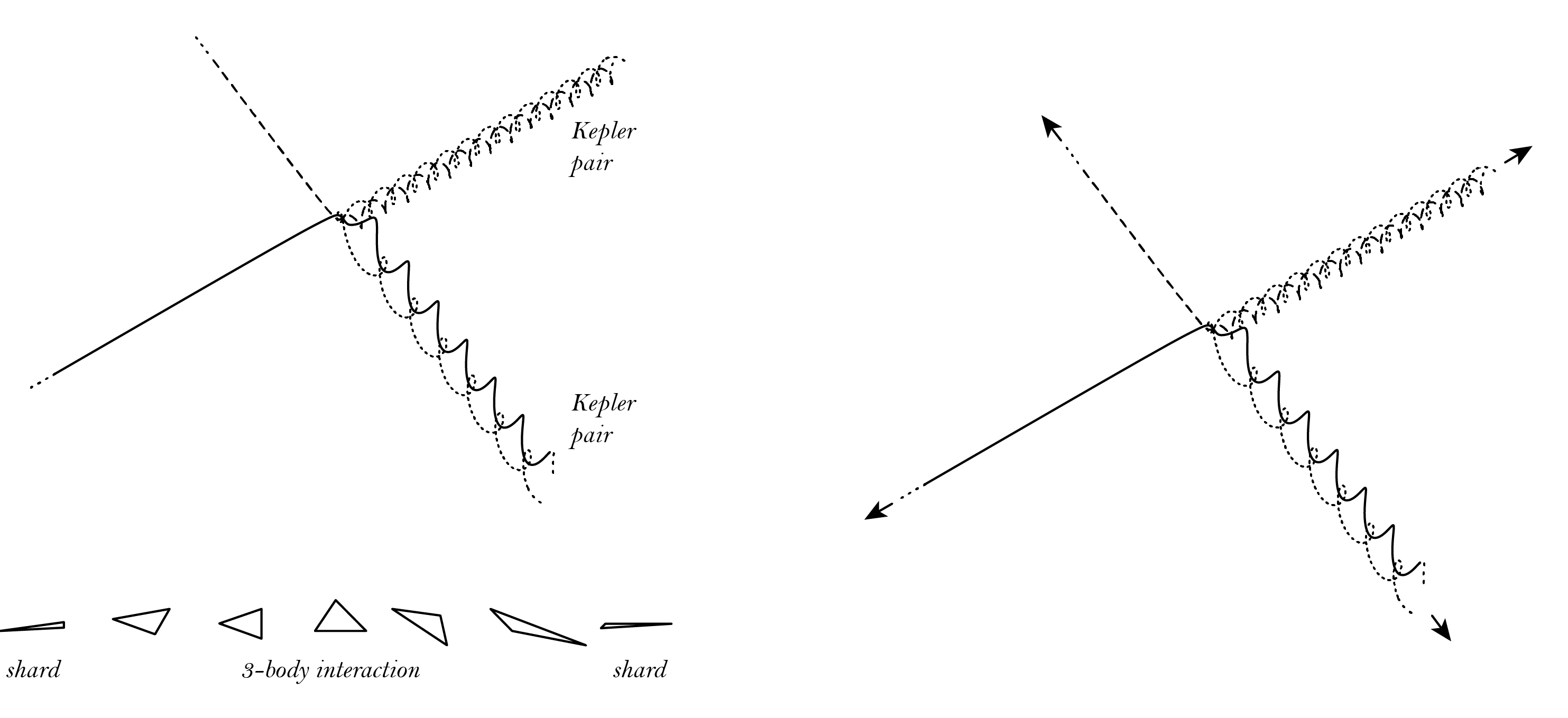}
\caption{\small The \tb solution described in the text. The description in the text applies to each of the two diagonals, which intersect in the figure at the Janus point.
\normalsize} \label{fig}
\end{center}
\end{figure}

For my purposes it will be sufficient to explain how models of rods, clocks, and compasses emerge already in the simplest \mb model universe---the \tb with vanishing energy and angular momentum. In accordance with case 3 as applied to the Newton potential in Sec.~\ref{bda}, it follows that the generic solutions have a Janus point. As detailed in \cite{jan}, chapter 7, the motion of the three particles, which is necessarily planar because the angular momentum is zero, is chaotic in the neighbourhood of the Janus point but in both directions away from it the system in all but a zero-measure set of solutions breaks up into a single particle, a singleton, and a pair that flies off the other way and settles down into ever better Keplerian motion around the pair's centre of mass. Asymptotically, there is therefore the singleton moving inertially in one direction and a Kepler pair with centre of mass moving inertially in the opposite direction with each of its particles having the same fixed direction of its major axis, which therefore defines a compass direction. The pair becomes a clock that `ticks' as each period is completed, while its major axis stablises to become a rod. Asymptotically, the three-body system becomes a universe that extends from the singleton to the pair centre of mass with a size that increases as measured by the emergent Kepler-pair rod. In the illustration in Fig.~\ref{fig} the triangles shown on the left have, since there is no external ruler, purely nominal sizes.

There are therefore two ways in which the universe can be said to expand: in a Newtonian representation in which such a solution is obtained before projection to shape space and intrinsically in the sense just described. It is only asymptotically, as the clocks get ever better, that the two rates of expansion agree. Expansion in the first sense relies on the implicit absolute scale, which I therefore find problematic. The intrinsic expansion obtained in the second sense matches the practice in observational cosmology. Surely the interpretation of the associated bidirectional arrows of time should be based on it. I may mention that, as explained in \cite{jan}, chapter 12, the emergence from chaos of rods and clocks that measure intrinsic expansion is universal for non-negative energy and is even more striking for $N \gg 3$ than in the \tbn.

\section{The Two Controlling Quantities\label{tcf}}

\begin{figure}
\begin{center}
\includegraphics[width=0.35\textwidth]{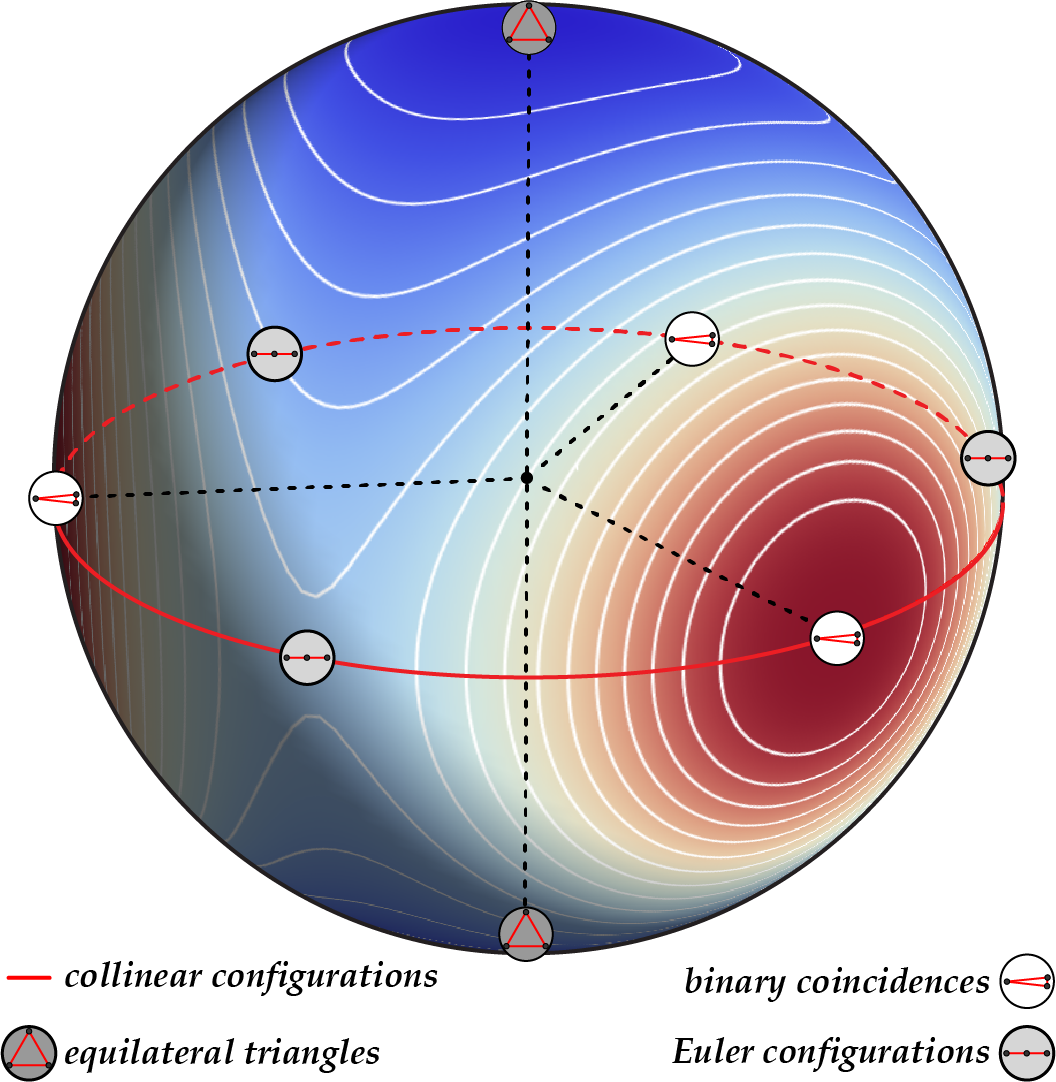}
\caption{\small The three-body  shape sphere with contours of the shape potential $V_\textrm{\scriptsize shape}$. 
\normalsize} \label{figs}
\end{center}
\end{figure}

By definition, shape space \s is a dimensionless arena constructed using pure numbers. It follows that only dimensionless quantities can determine the paths followed by \n solutions in \sn. The Newton potential \isa cannot be one of them since it is dimensionful. It has long been recognised in the \mb community that what really counts is \isa made dimensionless through multiplication by $\sqrt{I_\textrm{\scriptsize cm}}$ (\ref{icm}). It is then called the \e{normalised Newton potential} or, suggestively, the \e{shape potential} $V_\textrm{\scriptsize shape}$:
\be
V_\textrm{\scriptsize shape}:=\sqrt{I_\textrm{\scriptsize cm}}V_\textrm{\scriptsize New}.\label{spot}
\ee
In \cite{bkl,entaxy,jan} the negative of $V_\textrm{\scriptsize shape}$, denoted by $C$, is called the \e{complexity} because it is also a measure of the extent to which a distribution of mass points is structured or clustered as opposed to being uniformly distributed.\footnote{At the end of chapter 18 in \cite{jan} I make the tentative suggestion that in the (Newtonian) quantum universe the complexity literally is time. If this is correct, the strange role of the anomaly (\ref{ano}) emerges with the classical universe.}

For the \tb \s has $3\times 3-7=2$ dimensions and the system's possible shapes can be represented by points on a sphere as in Fig.~\ref{figs}.  
Chiral triangles have the same longitude and opposite latitudes; degenerate collinear shapes lie on the equator. The binary coincidences correspond to shapes in which the ratio of the separation between two of the particles divided by the distance to the third tends to zero. Besides these three singular points there are three extrema (saddles) of \np known as Euler configurations. The absolute minimum of $C$, the negative of \npn, is always at the equilateral triangle and in the case of equal masses shown here is located at the poles. The contours of the complexity show that $C$ has infinitely high peaks at the binary coincidences. The complexity is the first of the two controlling quantities that I discuss in this section.

\begin{figure}
\begin{center}
\includegraphics[width=0.35\textwidth]{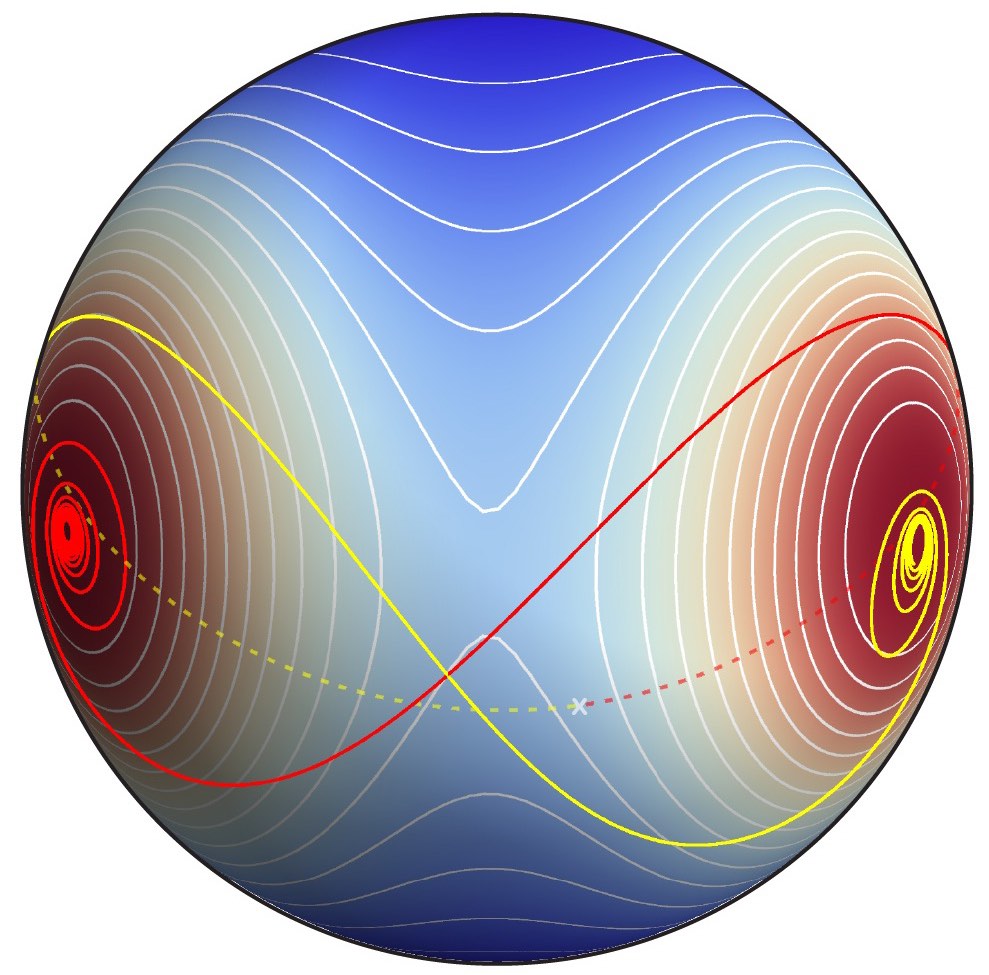}
\caption{\small The \tb solution described in the text with Janus point at the white cross on the `back' of the shape sphere. 
\normalsize} \label{fign}
\end{center}
\end{figure}

The other is identified in the Machian relational setting \cite{jan}, in which the energy and angular momentum are exactly zero. Then the degrees of freedom are $3N-7$ numbers that determine the shape, $3N-8$ that determine a two-sided direction in \sn, and the single extra variable (\ref{ano}) that I have called the anomaly.\footnote{I call it the \e{creation measure} in \cite{jan}, chapter 8, because it causes the solution curves in \s to deviate from geodesic paths and tend to shapes that are progressively more structured, the simplest examples of which are the `Kepler-pair--singleton universes' of Sec.~\ref{ue}.\label{crm}} A solution is determined by the initial point and direction in \s together with a value of $a$, which becomes an epoch-dependent variable along with the $6N-15$ shape dofs and measures the bending of the solution curve away from geodesic motion. The evolution is solely determined by the reaction of the instantaneous shape dofs and $a$ to the shape potential (\ref{spot}) just as in the conventional representation the state reacts to the gradient of \isan. However, as noted in \cite{bkl} and \cite{jan}, chapter 11, the evolution in \s is not symplectic (Hamiltonian) but non-symplectic with an effective friction term. The conventional interpretation of this is that projection to shape phase space of symplectic dynamics with scale makes it non-symplectic. But if, following \cite{dave}, one starts from what can actually be observed---shapes---it is the introduction of an extended phase space with scale that symplectifies the physical non-symplectic dynamics. The apparent effect of friction can be seen in Fig.~\ref{fign}, in which the Janus point is at the white cross at the `back' of the shape sphere and the unparametrised and undirected curve is more or less geodesic until it gets near the binary coincidences (one is on the `back' of the shape sphere), where it is trapped and spirals around the complexity peaks, reflecting the creation of Kepler pairs and singletons.

As noted at the end of the previous section, there are two ways in which one and the same solution can be interpreted. The traditional one is in the familiar Newtonian framework in which time and scale are absolute and Einstein's `sin' is committed: the existence of external rods and clocks that measure them are presupposed. In this interpretation the universe expands absolutely in both directions away from the Janus point. The concomitant change of shape during this process plays no role in the conclusion that the universe expands. Moreover, as in \cite{dag} and the purely inertial model of \cite{car,jo}, the sole criterion used to identify bidirectional arrows of time is the extrinsic (absolute) expansion present in the Newtonian representation.

The alternative, for which I argue, is the objective interpretation in terms of ratios, which can be read off directly from \sn, in this case the shape sphere, as shapes defined by pure numbers (angles). In this interpretation, absolute expansion can play no role but there is intrinsic expansion as the triangle shapes become evermore shardlike as in Fig.~\ref{fig}. Just as the astronomical unit defines the first leg of the distance ladder in cosmology, the Kepler-pair major axis defines the unit that measures the distance to the singleton and the current observable size of the three-body universe. Bidirectional arrows are defined within the timeless solution curve in Fig.~\ref{fign} through the fact that, once the solution curve is sufficiently far from the Janus point, there is a clear tendency for it to ascend ever higher the complexity peaks though not monotonically, a point to which I will return below. The only thing I want to emphasise at this stage is that bidirectional arrows are defined intrinsically and, as already noted, in line with the practice of observational cosmologists.

\section{Entropy and Confined and Unconfined Systems\label{dis}}

Clausius took the decisive step in the discovery of thermodynamics in 1850 by eliminating the notion of caloric in Carnot's theory of steam engines, which cannot function unless the steam is confined in a cylinder, or box. Clausius's subsequent discovery of entropy and the entire development of thermodynamics and statistical mechanics were made accordingly in the context of confined systems, which permit reversible transformations from one equilibrium state to another. In its most abstract form, statistical mechanics became the study of dynamical systems with a compact phase space and, accordingly, a bounded Liouville measure. In his definitive study \cite{gibbs}, Gibbs said the system must not expand to fill an infinite space and the momenta must satisfy a certain bound. My interest is the limit on the spatial extent, in connection with which Gibbs commented that in an infinite space thermodynamic equilibration to maximum entropy is not possible.

In a system whose phase space has bounded Liouville measure {\pon}'s recurrence theorem holds, and is the reason why Boltzmann, in his famous debate in the 1890s with Zermelo \cite{jan, brush} was, like everyone since, unable to identify the ultimate origin of the arrow of time though he did reach the now widely accepted view that entropy increase determines the experienced direction of time. It is interesting that Boltzmann did comment, though only in a footnote, that the recurrence theorem might not hold for the universe in its totality, in particular should it have infinite extent. In this connection note that a confined system whose size is increased reversibly has a well-defined and increasing entropy precisely because, as Gibbs noted, the container permits equilibration and a larger volume corresponds to a greater entropy. In statistical mechanics a formal box is also needed to define a fine-graining scale for the count of a macrostate's microstates. Without such a box an epistemologically suspect absolute scale must be assumed. For this reason, I question whether entropy can be defined sensibly in unconfined systems and think the bidirectional arrows discussed in \cite{dag,car,jo} call for a different interpretation (though as in \cite{bkl,entaxy,jan} I certainly agree they do suggest a past hypothesis may not be needed to explain the arrow of time).\footnote{The existence of bidirectional arrows of time has hardly been noted in the literature even though Lagrange discovered them in the unconfined \tb in 1772. To my knowledge such arrows, though with origin unrelated to the Lagrange--Jacobi relation (\ref{lj}) and with no clear dynamical underpinning, first appeared in the arrow-of-time literature in \cite{sak,ag,cc,et}. In all these cases, as in \cite{dag}, they are said to be entropic and thus arrows of increasing disorder. The existence of such arrows in the \nb were noted by my collaborators and myself in \cite{bkl} and discussed in a more comprehensive setting in \cite{entaxy,jan}. The `Janus cosmological model' \cite{jpp}, inspired in part by Sakharov's suggestion in \cite{sak} that bidirectional cosmological entropic arrows \e{could} exist and if so have interesting implications, shares Janus in its name with \cite{entaxy,jan} but little else that I can see.} 

A comment that, near the end of his life, Einstein made about thermodynamics is here in place \cite{auto}: ``It is the only physical theory of universal content which I am convinced that, within the framework of applicability of its basic concepts, will never be overthrown.'' I quote it for the caveat: does the universe come within the framework of applicability? Thermodynamics was discovered eight decades before expansion of the universe. Are we sure it has a phase space of bounded measure? Is it in a box? Dynamical systems can be divided into those that have a bounded phase space and those that do not. As far as thermodynamics and statistical mechanics are concerned, they only apply rigorously to the former. But strictly speaking, no dynamical systems with bounded phase space exist anywhere in the universe. Thermodynamics has its great universality because innumerable systems which are---spatially and temporally---isolated to lesser or greater accuracy do exist even though none exist forever.

To see the difference between confined and unconfined systems, consider the three-body solution described above.\footnote{For a more extensive discussion, see \cite{entaxy}, in which my collaborators and I developed more fully the implications of the existence of Janus-point solutions and introduced the notion of \e{entaxy}. At that stage, while we did know of their existence, we did not explore the implications of the zero-measure  `total-explosion' solutions that exist in accordance with the fifth possibility listed in Sec.~\ref{bda}. It was only from about 2018 that we began, for the reasons explained in chapters 16--18 in \cite{jan}, to take their vanishing size seriously as \n models of the big-bang in general relativity. In such solutions, the formation of branch systems (about to be discussed in the main text) will accompany growth of the complexity as happens in the Janus-point solutions considered here, but I think the special initial state of the `total-explosion' solutions makes the notion of entaxy have questionable value for them; an account in terms of complexity growth accompanied by the formation of branch systems is adequate. Moreover, for the reasons presented in chapter 19 of \cite{jan}, I become increasingly confident that, valuable as the entropy concept is for effectively confined systems at or near equilibrium, the better characterisation of what actually happens in the universe is captured by Kelvin's 1852 notion of dissipation of mechanical energy, though, again as argued in the same chapter in \cite{jan}, I would use the expression `spreading of energy'. Kelvin argued that \e{stores} of mechanical energy are subject to dissipation, but he had no explanation of their origin. I believe the account in \cite{entaxy} and chapter 19 of \cite{jan} does outline the closing of that lacuna.\label{lac}} In it, the behaviour in the vicinity of the Janus point is chaotic and unpredictable, but the invariable asymptotic formation of Kepler pairs as structures that are rods, clocks and compasses all in one must surely be interpreted as the creation of order, not disorder. Now suppose such a system in a box and an initial condition with the particles near its centre. There may well be an initial tendency for the system to break up into a pair and singleton, but as soon as the particles hit the walls of the box and bounce off it elastically all possibility of Kepler-pair formation is eliminated.

This underlines the difference from coventional statistical mechanics, in which macro- and microstates in a bounded phase space are widely considered. The possible microstates are associated with \e{microhistories}. As Zermelo pointed out to Boltzmann's discomfort, they must all return infinitely often arbitrarily close to any state through which they had previously passed. If the system is in addition ergodic, all states will be visited at some stage. However, in the \mb model considered in Sec.~\ref{bda} such behaviour is impossible since the dilatational momentum $D$ (\ref{dil}) is a Lyapunov variable; its existence rules out recurrence or ergodicity and puts a question mark over any attempt to characterise systems with unbounded phase spaces in conventional entropic terms. 

If recurrence is the hallmark of isolated systems with bounded phase space, its lack in the \mb solutions that we have been considering does not prevent the formation within them of the \e{branch systems} just mentioned in footnote~\ref{lac} and discussed in detail in \cite{entaxy}. This is exactly what happens in the three-body solution described above. In it the Kepler pair settles down into behaviour that, like the complete solution, satisfies the Lagrange--Jacobi relation (\ref{lj}) but with negative energy. In accordance with the virial theorem, which is derived from (\ref{lj}), this allows the pair to tend asymptotically to possession of a time-independent virial.\footnote{The marked difference between the behaviour of the three-body system as a whole and the Kepler pair is that, despite both satisfying (\ref{lj}), the former has positive energy and a Lyapunov variable while the latter has negative energy, which is what allows it to self-confine. This is also what allows more or less isolated subsystems to form and, before decaying, virialise approximately in all \mb systems with non-negative energy.} A virtually constant virial is precisely what characterises the behaviour of an equilibrated isolated system. Thus, in a universe subject to a Lyapunov variable systems that come into existence out of equilibrium can subsequently equilibrate. It should be said though that a Kepler pair is special. Virtually all gravitational systems that self-confine do evaporate, globular clusters for example over a time of the order of the age of the universe \cite{hh} and, in accordance with Hawking's famous result, black holes over a vastly longer time.

\begin{figure}
\begin{center}
\includegraphics[width=0.8\textwidth]{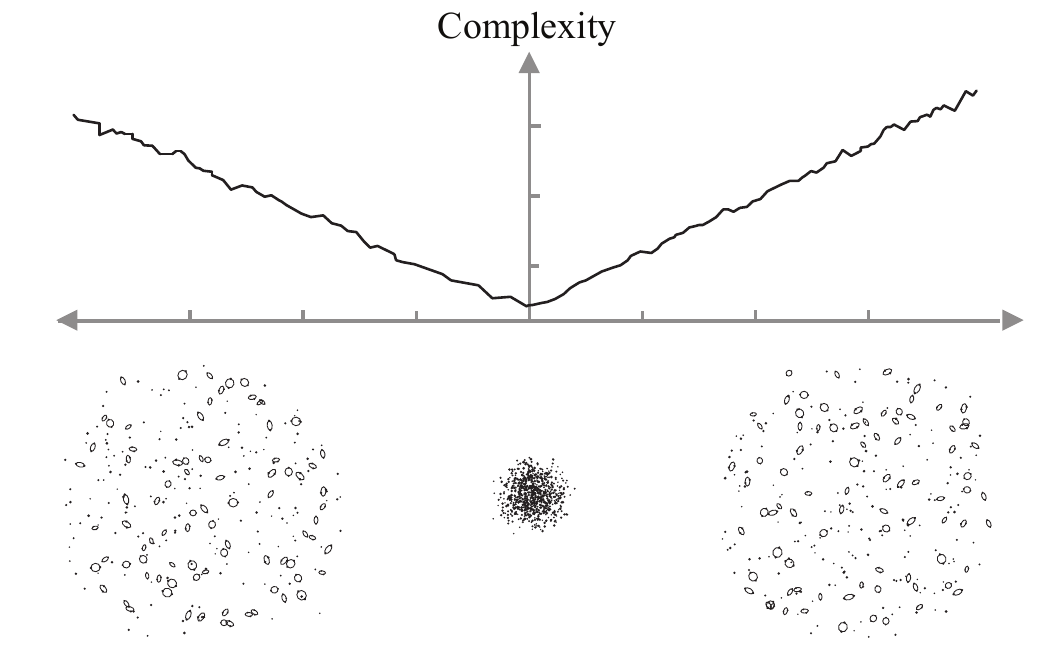}
\caption{\small Growth of the complexity either side of the Janus point as found numerically by Jerome Barkley for 1000 equal-mass particles and `artist's impression' of the system's appearance: very uniform in the region of the Janus point and clustered either side of it. The figure is reproduced from \cite{bkl}.
\normalsize} \label{comp}
\end{center}
\end{figure}

To come to a conclusion, there is a clear alternative to an implicit absolute scale in the description and interpretation of unconfined systems proposed as model universes. It is to represent their essential content as proposed in \cite{bkl,entaxy,jan} in scale-invariant terms in shape space \sn. Without any loss of physical information, this eliminates all questionable features introduced by Newton's absolutes and brings out clearly what is actually happening: the complexity $C$, a pure number and invariant of the similarity group, has a clear tendency to increase (as seen in Fig.~\ref{comp}), admittedly and like entropy with small fluctations in systems with many particles, either side of the Janus point. Moreover, the increase corresponds to growth of order, not disorder. For although the Kepler pair `equilibrates' it still, through being a rod, clock, and compass all in one, represents remarkable order, within which the singleton moves in an inertial frame in accordance with Newton's first law, also with ever increasing accuracy. See also the account in chapter 12 of \cite{jan} of the asymptotic order, already mentioned at the end of Sec.~\ref{ue}, that is created for arbitrarily large $N$. 

The antithesis of behaviours described in this paper matches the antithesis of defining conditions: confined, recurrent, and equilibrating in one; unconfined, non-recurrent, and non-equilibrating in the other. The broad characterisation of the typical unconfined behaviour like that seen in Fig.~\ref{comp} is of one from \e{uniformity} of the particle distribution in the vicinity of the Janus point to ever increasingly structured \e{variety} with increasing distance from it. This effect is especially pronounced in the total-explosion \mb solutions (case 5 in Sec.~\ref{bda}) that commence with zero size and are described in chapter 16 of \cite{jan}. I suggest there that such behaviour may hint at the form of the quantum law that governs the universe.

The fact I want to emphasise most is that all physical facts come to us as ratios. This puts heat death of the universe in a different perspective. In frequently found modern descriptions, it is no longer argued that all temperature differences are eliminated but only that, as the universe expands, energy densities decrease to such an extent that life becomes impossible. But this is an account in an \e{extended} phase space with implicit absolute units. Since only ratios are physical, they can still have all possible values and be experienced by real beings in the actual shape-changing universe. Evolution of the universe need not cease. Although we cannot expect to be there to see what happens, there is no reason to suppose that nothing will.

\vspace{.05in}

{\flushleft{\bf Acknowledgements.} My thanks above all to Tim Koslowski, Flavio Mercati, and David Sloan for their input.}

\end{document}